\documentclass[screen]{acmart}

\usepackage{acmart-taps}
\usepackage{graphicx,booktabs,tabularx,mathtools,natbib,enumitem}
\usepackage[capitalise,nameinlink,compress]{cleveref}
\usepackage{tikz,xcolor}

\renewcommand{\quote}[1]{\textit{``#1''}}

\definecolor{thinkaloud1}{HTML}{3D5A80}
\definecolor{focusgroups}{HTML}{D7A751}
\definecolor{thinkaloud2}{HTML}{00897B}
\definecolor{grey}{HTML}{7f7f7f}

\def\A#1{\textcolor{thinkaloud1}{\textbf{A#1}}}
\def\B#1{\textcolor{focusgroups}{\textbf{B#1}}}
\def\C#1{\textcolor{thinkaloud2}{\textbf{C#1}}}
\def\G#1{\textcolor{focusgroups}{\textbf{G#1}}}

\DeclareRobustCommand\circledsmall[1]{\tikz[baseline=(char.base)]{
    \node[shape=circle,draw,inner sep=1pt,color=white,fill=grey] (char) {#1};}}

\newcommand{\change}[1]{#1}
\newcommand{\revision}[1]{#1}

\copyrightyear{2025}
\acmYear{2025}
\setcopyright{rightsretained}
\acmConference[LAK 2025]{LAK25: The 15th International Learning Analytics and Knowledge Conference}{March 03--07, 2025}{Dublin, Ireland}
\acmBooktitle{LAK25: The 15th International Learning Analytics and Knowledge Conference (LAK 2025), March 03--07, 2025, Dublin, Ireland}
\acmDOI{10.1145/3706468.3706470}
\acmISBN{979-8-4007-0701-8/25/03}

\begin{document}

\title{Designing Visual Explanations and Learner Controls to Engage Adolescents in AI-Supported Exercise Selection}

\author{Jeroen Ooge}
\orcid{0000-0001-9820-7656}
\affiliation{%
  \institution{Utrecht University}
  \department{Department of Information and Computing Sciences}
  \streetaddress{Princetonplein 5}
  \city{Utrecht}
  \postcode{3584 CC}
  \country{The Netherlands}
}
\affiliation{%
  \institution{KU Leuven}
  \department{Department of Computer Science}
  \streetaddress{Celestijnenlaan 200A}
  \city{Leuven}
  \postcode{3001}
  \country{Belgium}
}
\email{j.ooge@uu.nl}

\author{Arno Vanneste}
\orcid{0009-0003-5164-1107}
\affiliation{%
  \institution{KU Leuven}
  \department{Department of Computer Science}
  \streetaddress{Celestijnenlaan 200A}
  \city{Leuven}
  \postcode{3001}
  \country{Belgium}
}
\email{arnovanneste96@gmail.com}

\author{Maxwell Szymanski}
\orcid{0000-0002-6506-3198}
\affiliation{%
  \institution{KU Leuven}
  \department{Department of Computer Science}
  \streetaddress{Celestijnenlaan 200A}
  \city{Leuven}
  \postcode{3001}
  \country{Belgium}
}
\email{maxwell.szymanski@kuleuven.be}

\author{Katrien Verbert}
\orcid{0000-0001-6699-7710}
\affiliation{%
  \institution{KU Leuven}
  \department{Department of Computer Science}
  \streetaddress{Celestijnenlaan 200A}
  \city{Leuven}
  \postcode{3001}
  \country{Belgium}
}
\email{katrien.verbert@kuleuven.be}

\renewcommand{\shortauthors}{Ooge et al.}

\begin{abstract}
E-learning platforms that personalise content selection with AI are often criticised for lacking transparency and controllability. Researchers have therefore proposed solutions such as open learner models and letting learners select from ranked recommendations, which engage learners before or after the AI-supported selection process. However, little research has explored how learners -- especially adolescents -- could engage during such AI-supported decision-making. To address this open challenge, we iteratively designed and implemented a control mechanism that enables learners to steer the difficulty of AI-compiled exercise series before practice, while interactively analysing their control's impact in a \textit{what-if} visualisation. We evaluated our prototypes through four qualitative studies involving adolescents, teachers, EdTech professionals, and pedagogical experts, focusing on different types of visual explanations for recommendations. Our findings suggest that \textit{why} explanations do not always meet the explainability needs of young learners but can benefit teachers. Additionally, \textit{what-if} explanations were well-received for their potential to boost motivation. Overall, our work illustrates how combining learner control and visual explanations can be operationalised on e-learning platforms for adolescents. \revision{Future research can build upon our designs for \textit{why} and \textit{what-if} explanations and verify our preliminary findings.}
\end{abstract}



\maketitle

\section{Introduction}
Education is increasingly adopting AI-supported technologies to personalise learning~\cite{verbert2012contextaware}, and shifting from traditional classrooms to e-learning environments~\cite{salau2022stateoftheart}. Evidence grows that AI-supported technologies can assist in improving learning, yet there remain concerns about their transparency and controllability because the applied AI techniques are not always inherently interpretable. Interestingly, education has been studying these aspects since the 1990s. For example, open learner models were proposed to foster metacognition and self-regulation by showing learners what educational systems know about them, including inferred skill mastery, preferences, and progress~\cite{rahdari2020personalizing,bull2004open,bull2007student,bull2020there,bodily2018open,conati2018ai}. These open learner models could furthermore serve as a starting point to let learners steer or negotiate their learner models~\cite{bull1995did,mabbott2006student}, thus operationalising learner control in AI-supported education. Other examples of such control include changing how learning materials are presented or selected, for instance, by choosing the desired difficulty~\cite{kay2001learner,brusilovsky2023ai,papousek2017should}.

Research into transparency and controllability is nowadays revived in both education and other AI application domains under the umbrella of explainable AI (XAI)~\cite{khosravi2022explainable,gunning2019darpa} and human-AI interaction~\cite{vanberkel2021humanai}. The XAI community has developed many different types of post-hoc explanations aimed to let people understand how AI models obtain their outcomes and behave under changes~\cite{guidotti2019survey}. For example, \textit{why} explanations clarify why a given prediction was being made and which features led to it, and \textit{what-if} explanations show how predictions would change under different inputs~\cite{liao2020questioning,lim2009why}. Importantly, developing effective explanations is as much of a design challenge as an algorithmic challenge because the effectiveness lies in the perception of the people receiving the explanations~\cite{liao2022humancentered}. Therefore, human-centred XAI focuses on how to build upon knowledge from the social sciences~\cite{miller2019explanation} and how to tailor explanations to different target audiences, for example, based on their background knowledge~\cite{mohseni2021multidisciplinary} or explainability needs~\cite{liao2020questioning}.

Despite the rich literature on transparency and controllability for AI, several aspects remain underexplored in education. Regarding transparency, it is unclear which of the many explanation types developed by XAI researchers are suitable for adolescents in an educational context: existing XAI studies mainly involve adult participants in application domains such as e-commerce and media. Therefore, user studies should assess whether adolescents need, understand, and benefit from explanations in AI-supported educational systems. Moreover, current explanations for AI outcomes typically have a static format~\cite{abdul2018trends} and fail to cognitively engage people in AI-supported decision processes~\cite{liao2022humancentered,miller2023explainable}. It is an open question whether interactive explanations can ameliorate this issue. Regarding controllability, there is little research on learner control mechanisms for selecting learning materials in collaboration with AI models~\cite{brusilovsky2023ai}. One possible reason is that learners are often assumed to have too little prior knowledge to exercise control over learning materials, especially when they are young~\cite{brusilovsky2023ai}. Yet, learner control has generally been considered motivating and enjoyable~\cite{long2017enhancing,clark2011elearning}, so further exploring how learners can steer AI models in educational systems seems worthwhile. Finally, studies that propose transparency or learner control solutions often lack or do not describe an elaborate design process involving multiple stakeholders. Doing so can better ground design choices, clarify the non-triviality of conducting such a design process, and give deeper insights into the advantages and disadvantages of different designs.

To address the above challenges, we conducted a four-stage design process with adolescents, teachers, EdTech professionals, and pedagogical experts, focusing on how \textit{why} and \textit{what-if} explanations, and learner control can be integrated on e-learning platforms that recommend exercises. These were our research questions:
\aptLtoX{\begin{itemize}
  \item \textbf{RQ1}. How can \textit{why} and \textit{what-if} explanations be operationalised on exercise-recommending e-learning platforms while meeting the explainability needs of adolescent learners and teachers?
  \item \textbf{RQ2}. How can adolescent learners steer the difficulty of recommended exercises on e-learning platforms in line with their needs for control and the appropriateness of such control?
\end{itemize}}{\begin{itemize}[leftmargin=*]
  \item \textbf{RQ1}. How can \textit{why} and \textit{what-if} explanations be operationalised on exercise-recommending e-learning platforms while meeting the explainability needs of adolescent learners and teachers?
  \item \textbf{RQ2}. How can adolescent learners steer the difficulty of recommended exercises on e-learning platforms in line with their needs for control and the appropriateness of such control?
\end{itemize}}

Our extensive design process resulted in a pre-practise control interface wherein learners can indicate their preferred difficulty for the upcoming AI-composed exercise series while seeing either a corresponding \textit{what-if} explanation or motivational feedback. During group discussions, think-aloud sessions, and focus groups, we learned that \textit{why} explanations do not necessarily meet adolescents' needs, but can benefit teachers in multiple ways. Furthermore, \textit{what-if} explanations hold the potential to boost learners' motivation, thus manifesting a so far unexplored bridge between XAI and motivation. We hope our preliminary findings inspire follow-up studies in education that explore the interplay between transparency and learner control and their impact on metacognition and motivation.

\section{Background and Related Work}
\label{sec:background}

\subsection{Visual Explanations for AI Models}
Researchers focused on algorithmic XAI have developed many post-hoc explanation techniques to provide insights into how black-box AI models obtain their outcomes and behave under parameter or input changes~\cite{adadi2018peeking,barredoarrieta2020explainable,guidotti2019survey,stiglic2020interpretability,vilone2020explainable,du2019techniques,montavon2018methods,zhang2020explainable}. As these explanations typically capture a lot of information, visualisations are often applied for effective communication. Examples include visualising feature importances~\cite{bertrand2023questioning,lundberg2017unified}, interactive sensitivity analysis~\cite{szymanski2021visual,hohman2019gamut}, \textit{why} explanations about recommendation processes~\cite{bostandjiev2012tasteweights}, and example-based explanations~\cite{cai2019effects}. For education in specific, \citet{ooge2022explaining} justified recommended exercises on an e-learning platform by visualising information about the collaborative filtering step and found this increased initial trust in the platform. Furthermore, \citeauthor{barria-pineda2018finegrained}~\cite{barria-pineda2018finegrained,barria-pineda2019making} justified recommended exercises by showing how likely learners are to solve them, and \citet{abdi2020complementing} complemented recommendations with a visual open learner model. However, few of these studies targeted adolescents, conducted rigorous explainability needs assessments, and involved multiple educational stakeholders in the design process of explanations~\cite{conati2021personalized}.

\subsection{Control Over AI Models}
The rise of AI-supported systems has raised questions about how control over decision-making should be distributed among AI systems and the people using them: should AI systems be given full control, or should there be human-AI collaboration~\cite{vanberkel2021humanai}? This leads to more questions such as when people should exert control, how control mechanisms should be designed, and what the role of personal characteristics is. Furthermore, even though explanations and control mechanisms have different goals, they can be deemed ``two sides of the same coin''~\cite{storms2022transparency}: having control over an AI model can grow better understanding of how it behaves and makes decisions, and, conversely, seeing how an AI model obtains its outcomes might evoke a higher need for correcting or steering it. This shows why control over AI is tightly linked to explainable AI and has been extensively studied in the general setting of recommender systems~\cite{jannach2017user}. In educational contexts, many control mechanisms have been proposed~\cite{brown2016learner} under the umbrella of \textit{learner control} to actively involve learners in their learning process and stimulate metacognition~\cite{brusilovsky2023ai,kay2001learner,kay2019data}. Examples include: directly changing or negotiating about the learner model inferred by AI~\cite{mabbott2006student,bull1995did}, and selecting the difficulty level of learning materials~\cite{long2016masteryoriented,long2017enhancing,papousek2017should}.

\section{Methods and Materials}
\begin{figure*}
  \includegraphics[width=.9\linewidth]{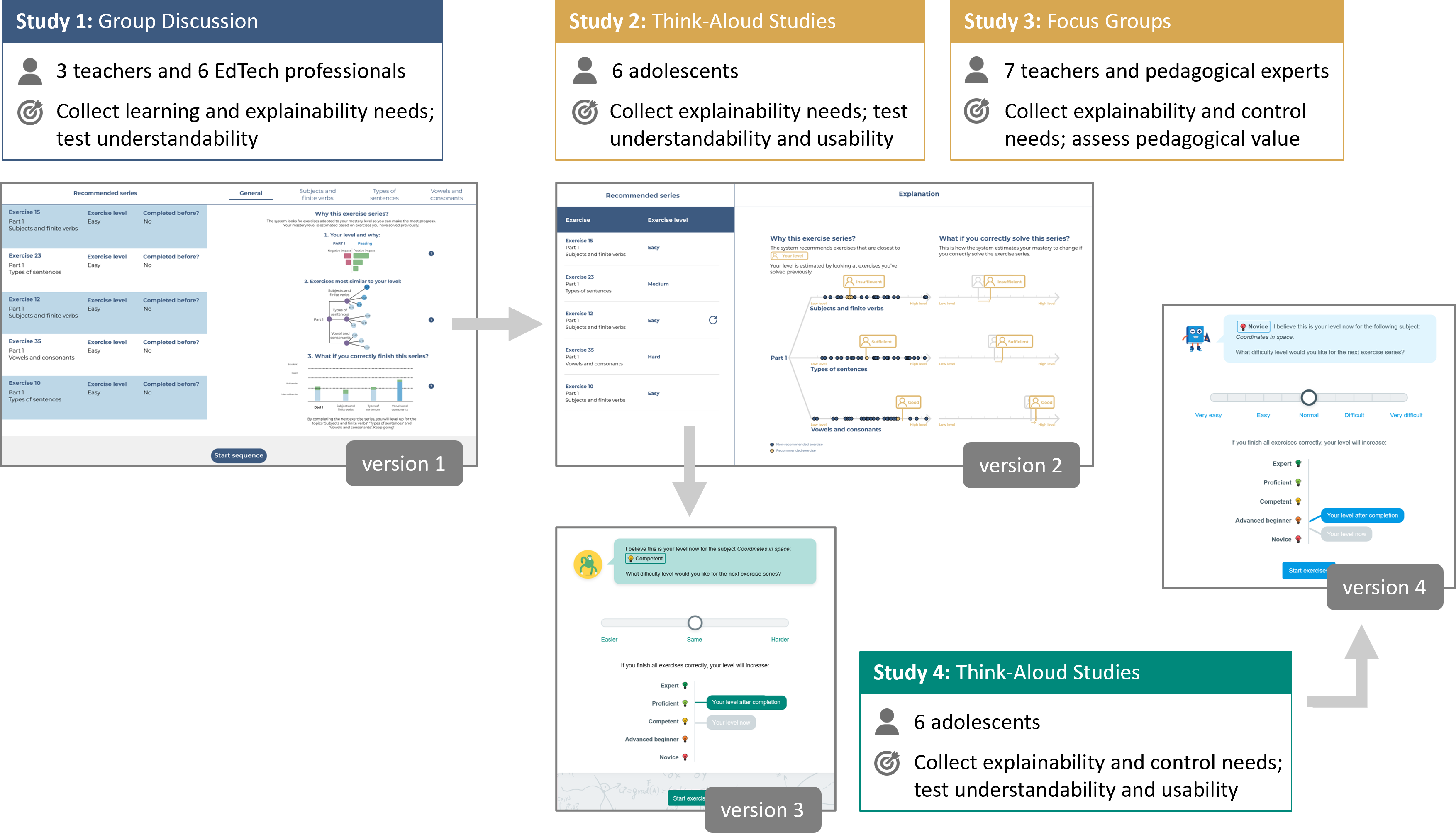}
  \caption{Overview of the four user studies we conducted with adolescents, teachers, EdTech professionals, and pedagogical experts, together with the goals of each study. Larger screenshots of our interfaces are presented in Figures~\ref{fig:prototype1} to \ref{fig:flow}.}
  \Description{Four screenshots representing our design iterations. The first screenshot is linked to study 1, the second to studies 2 and 3, and the third to study 4.}
  \label{fig:overviewstudies}
\end{figure*}

Our research targeted middle and high school students and was positioned in the context of an e-learning platform that uses AI to compose series of recommended exercises in maths or languages. To answer our research questions, we consulted adolescents, teachers, EdTech professionals experienced with developing e-learning platforms, and pedagogical experts in an iterative design process including four user studies, as depicted in Figure~\ref{fig:overviewstudies}. Each study informed design changes in the next iteration and involved different participants recruited via schools and EdTech companies in \anon[REGION]{Belgium (Flanders)}. As prior XAI research in the context of K-12 education is sparse, multiple iterations with small sample sizes allowed us to refine our designs in a targeted way while collecting rich feedback from multiple educational stakeholders.

\textit{Study~1} acted as a pilot study, consisting of a 2-hour group discussion with 3 high school teachers and 6 EdTech professionals. We collected learning needs, gathered feedback on low-fidelity prototypes in Figma, and brainstormed about learner control and explanations on e-learning platforms that recommend exercises. In \textit{study~2}, we conducted think-aloud sessions with 6 adolescents after parental consent. In these studies of 15--30 minutes each, participants completed predetermined tasks in a Figma prototype while articulating their thoughts so we could observe potential usability issues. Follow-up questions assessed understanding of the explanations and explainability needs, and participants received a \revision{\anon[CURRENCY]{€15}} 
voucher. In \textit{study~3}, we guided two focus groups of 2 hours each with 7 teachers and pedagogical experts. Participants discussed how our explanation interfaces could be integrated into e-learning platforms, whether they met students' and teachers' needs, and how they may affect motivation, trust, and metacognition. Finally, \textit{study~4} encompassed 6 more think-aloud sessions with adolescents: we followed a protocol similar to study~2 but additionally focused on learner control and implemented our design on a fully functional e-learning platform in Drupal such that participants could experience its full workflow. To measure cognitive engagement, we observed participants' reactions while interacting with the interfaces (e.g., hesitation, reflection, use of the slider and explanations) and documented their thought process and decision-making.

All think-aloud sessions and focus groups were recorded for transcription and analysis after written informed consent. The analysis consisted of thematically grouping relevant quotes and combining them with our observations into main findings along with their prominence. We conducted all studies in \anon[LANGUAGE]{Dutch}, but translated quotes and interfaces for presentational purposes.

\section{Results}
\label{sec:designprocess}
Our design process involved four studies, depicted in Figure~\ref{fig:overviewstudies}. We present participants' attitudes towards our designs for explanations and learner control and describe how the designs evolved as a result.

\subsection{Study 1: Group Discussion}
\begin{figure*}
  \centering
  \includegraphics[width=.75\linewidth]{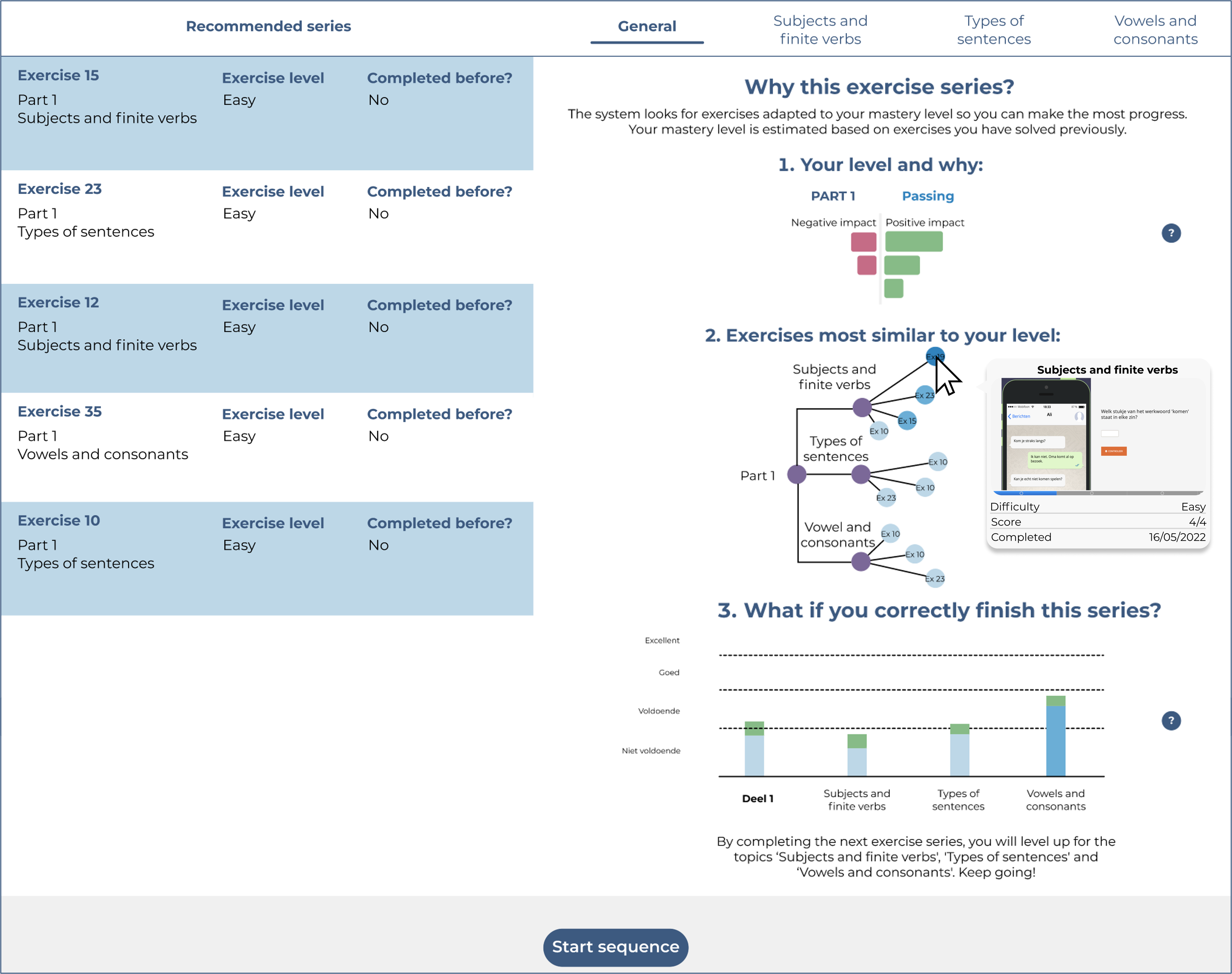}
  \caption{Screenshot of our first prototype. Left: the recommended exercise series. Right: an explanation panel shows how previous attempts established learners' mastery levels and \textit{why} and justifies recommendations with \textit{what-if} explanations.}
  \Description{Fully described in the caption and main text.}
  \label{fig:prototype1}
\end{figure*}

Based on our experience with explainable recommender systems in education \anon[REF]{\cite{ooge2023steering,ooge2022explaining}} and the related work in Section~\ref{sec:background}, we combined several explainability ideas in the interface in Figure~\ref{fig:prototype1}, which learners would see before starting an AI-composed exercise series. To enhance transparency, we indicated difficulty levels and previous attempts next to recommended exercises, and a split bar chart visualised how past exercises established learners' mastery level. To justify recommendations, a \textit{why} explanation showed how recommended exercises are close to learners' mastery level, and a \textit{what-if} explanation showed the change in mastery for all topics in case learners would solve the series correctly.

The teachers and EdTech professionals in our group discussion appreciated the idea of personalising adolescents' learning process with an e-learning platform, corresponding to previous research~\cite{ooge2023steering}. Yet, teachers added that \quote{mainly high-level students need asynchronous differentiation to be challenged}, and questioned whether e-learning platforms are best suited to support low-level students. Furthermore, all participants suggested clarifying the visualisations and restricting the amount of information in the interface. Finally, teachers suggested focusing on explanations that \quote{boost motivation because adolescents often give up on learning.} Due to this last remark, \textit{what-if} explanations were highlighted as the most valuable component in our interface.

\subsection{Study 2: Think-Aloud Sessions}
\label{sec:study2}
\begin{figure*}
  \centering
  \includegraphics[width=.8\linewidth]{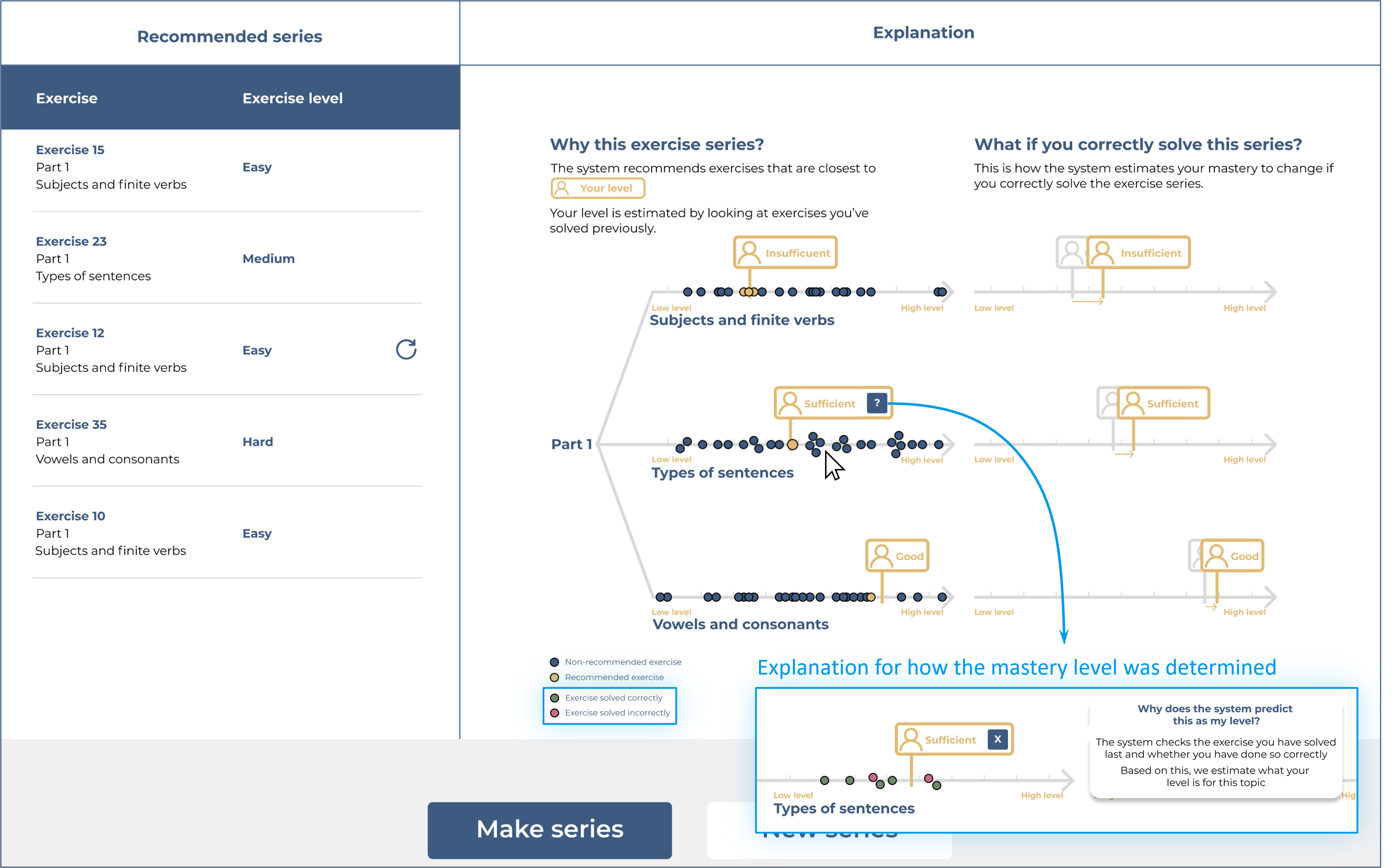}
  \caption{Screenshots of our second prototype with redesigned \textit{why} and \textit{what-if} explanations. Hovering scatter plots adds vertical jitter, and the info button in the mastery level labels reveals previous performance.}
  \Description{Fully described in the caption and main text.}
  \label{fig:prototype2}
\end{figure*}

Based on the findings in study~1, we simplified our designs as presented in Figure~\ref{fig:prototype2}. First, we streamlined the \textit{why} explanation: topics of recommended exercises were depicted as tree branches with all topic-specific exercises scattered over the branch in rising difficulty; learners' current mastery levels were indicated with a label. Highlights showed the recommended exercises lying close to learners' mastery, and learners could reveal previous attempts per topic. Furthermore, we made the \textit{what-if} explanation more prominent and visualised mastery level changes as shifting labels on axes identical to the tree branches.

To test our prototype's usability, assess whether explanations improve understanding of the recommendation algorithm, and elicit needs regarding personalised e-learning, we then conducted think-aloud sessions with 6~adolescents ({\textcolor{thinkaloud1}{\textbf{A1}}} and \A{2} in 8th grade, \A{3} in 9th grade, \A{4} and \A{5} in 10th grade, and \A{6} in 11th grade). Table~\ref{tab:thinkaloud1} summarises our findings, which we further distilled into the following three lessons.

\paragraph{Minimise Text, Tightly Integrate Explanations} Many participants did not carefully read annotations such as the legend and descriptions above the visualisations before we asked them to because \quote{it looked too serious} (\A{5}) or \quote{there was too much text} (\A{4}). The annotations could be more compact by merging the legend with the descriptions, similar to the consistently styled mastery labels. Still, \A{6} suggested that a brief tutorial explaining the visualisation could help. More importantly, even though only \A{1} mentioned it explicitly, we observed that participants did not link the explanations to the recommendations. This shows explanations should be tightly integrated with whatever they explain: simply presenting everything side-by-side might make adolescents ignore the explanations.

\paragraph{Inattentiveness Can Harm Mental Models} Although participants did not experience major usability issues, they initially found the many dots in the \textit{why} explanation overwhelming and needed time or clarification to grasp their meaning. Together with the previous observation of participants not carefully reading annotations, this led some participants to create inaccurate mental models by combining elements in the interface and prior expectations. For example, \A{1} wrongfully believed the recommendation algorithm applied collaborative filtering, assuming the bee swarm dots represented other learners whose mastery levels were compared to theirs. This illustrates how visual explanations causing cognitive overload can yield mental models that seem sensible but are wrong.

\paragraph{Adolescents Want Control and Motivation Supports} Finally, participants brought up two needs when using e-learning platforms. First, even though they saw many advantages in personalised recommendations, they also wanted to keep some degree of control over which exercises to solve. For example, \A{2}, \A{3}, and \A{6} suggested control in the form of re-making exercises when desired, potentially by re-purposing the scatter plot in the \textit{why} explanation as a way to select exercises. Alternatively, \A{1} and \A{5} proposed control in the form of hand-picking exercises from a list or table that indicates their difficulty level. Interestingly, participants also alluded to a need for motivation support. For example, \A{4} and \A{6} mentioned they found the \textit{what-if} explanation motivating as it showed a goal to work towards. Moreover, \A{1} appreciated a rationale for recommendations because \quote{if you immediately understand why [exercises are recommended] then you're more inclined to practise and come back to it.}

\begin{table*}
  \centering
  \footnotesize
  \caption{Main findings of our think-aloud sessions with the prototype in Figure~\ref{fig:prototype2}, ordered by overall theme and frequency.}
  \label{tab:thinkaloud1}
\begin{tabular}{@{}ll@{}}
	\toprule
	\textbf{Comment}                                                                                                        & \textbf{Participants}                    \\


	\midrule
	\multicolumn{2}{@{}l}{\textcolor{thinkaloud1}{\textbf{\textit{Why} explanation and understanding recommendations}}}                                                                                     \\
	Scatter plot is (initially) unclear (e.g., position and colouring of the dots are unclear)               & \A{1}, \A{2}, \A{3}, \A{4}, \A{5}, \A{6} \\
	Mastery labels are clear (e.g., because of the consistent lay-out in the text above the visualisation) & \A{1}, \A{2}, \A{3}, \A{4}, \A{5}, \A{6} \\
  Understands that recommendations are based on previous performance and mastery level                                           & \A{2}, \A{3}, \A{4}, \A{5}, \A{6}               \\
	Did not read text above the visualisation (e.g., because it was too small or too long)                                  & \A{1}, \A{4}                             \\
	Understanding why an exercise is recommended increases eagerness to solve it                                                & \A{1}                                    \\
  Wrongly believes recommendations rely on collaborative filtering                                                      & \A{1}                                    \\

	\midrule
	\multicolumn{2}{@{}l}{\textcolor{thinkaloud1}{\textbf{\textit{What-if} explanation}}}                                                                                 \\
	Clear that the visualisation shows progress in mastery                                                                         & \A{1}, \A{2}, \A{3}, \A{5}, \A{6}               \\
	Seeing progress is motivating                                                                                           & \A{4}, \A{6}                             \\
  
  \midrule
  \multicolumn{2}{@{}l}{\textcolor{thinkaloud1}{\textbf{Explanations overall}}}                                                                          \\
  Would not always look at explanations, but wants them on-demand (e.g., when making many mistakes)                                                          & \A{4}, \A{6}                                    \\
  Relation between recommendations and explanations is unclear                                                            & \A{1}                                    \\

	\midrule
	\multicolumn{2}{@{}l}{\textcolor{thinkaloud1}{\textbf{Control}}}                                                                                           \\
  Wants to re-make exercises (e.g., by clicking dots in the \textit{why} explanation)                                     & \A{2}, \A{3}, \A{6}                      \\
	Still wants to choose exercises (of a specific difficulty level) themselves                                             & \A{1}, \A{5}                             \\
	\bottomrule
\end{tabular}
\end{table*}

\subsection{Study 3: Focus Groups}
Because young learners may not always know what is best for their learning process, we also captured how teachers and pedagogical experts received our prototype in Figure~\ref{fig:prototype2}. Specifically, we conducted two focus groups (\G{1} and \G{2}) with respectively four (\B{1}--\B{4}) and three (\B{5}--\B{7}) participants. Table~\ref{tab:focusgroupsBackground} lists their background, and Table~\ref{tab:focusgroups} summarises the discussed themes. Overall, participants seconded several comments by \A{1}--\A{6} in Section~\ref{sec:study2} and gave insights into how teachers could benefit from explanations and how pedagogical practices could improve our designs.

\begin{table*}
  \centering
  \caption{Background of the teachers and pedagogical experts participating in our focus groups.}
  \label{tab:focusgroupsBackground}
\footnotesize
\begin{tabular}{@{}ll@{}}
  \toprule
  \textbf{ID} & \textbf{Gender and Background}\\
  \midrule
  \B{1} & {F} -- deputy director at high school, educational author, implemented ICT tools in college education\\
  \B{2} & {F} -- college researcher specialised in K-12 education, didactic support person for digital learning, former primary school teacher\\
  \B{3} & {F} -- scientific employee on educational technologies and partnering with industry, didactician, former K-12 educational author\\
  \B{4} & M -- educational support person and ICT coordinator at a college, former high school teacher\\
  \B{5} & M -- high school teacher of languages using ICT in class, pedagogical worker for special care education, educational author\\
  \B{6} & M -- manager at different educational publishers, former high school teacher of classical languages and history, pedagogue\\
  \B{7} & M -- product owner at K-12 educational publisher, former K-12 educational author, former primary school teacher\\
  \bottomrule
\end{tabular}
\end{table*}

\begin{table*}
  \footnotesize
  \caption{Main findings of our focus groups with teachers and pedagogical experts, ordered by overall theme and prominence. Participants discussed the prototype in Figure~\ref{fig:prototype2} with adolescents in mind.}
  \label{tab:focusgroups}
\begin{tabular}{@{}ll@{}}
	\toprule
	\textbf{Comment}                                                                                                                                                                                                                          & \textbf{Group} \\
	\midrule
	\multicolumn{2}{@{}l}{\textcolor{focusgroups}{\textbf{\textit{Why} explanation for students}}}                                                                                                                                                                \\
	Could be superfluous if students need to solve the exercises anyway                                                                                                                                                                       & \G{1}, \G{2}   \\
	Visually clean, but too complex and too detailed for adolescents                                                                                                                                                                             & \G{1}, \G{2}   \\
	Showing all exercises could give students the impression they need to solve them all                                                                                                                                                      & \G{1}          \\
	Only show exercises that fit students' mastery level or are useful to reach a goal (e.g., unlock levels with exercises)                                                                                                                   & \G{1}          \\
	Could be useful for students to choose exercises themselves                                                                                                                                                                               & \G{2}          \\

	\midrule
	\multicolumn{2}{@{}l}{\textcolor{focusgroups}{\textbf{\textit{Why} explanation for teachers}}}                                                                                                                                                                \\
	Useful for teachers to monitor students                                                                                                                                                                                                   & \G{1}, \G{2}   \\
	Suitable for teachers to understand how and why exercises are being recommended                                                                                                                                                           & \G{1}          \\
	Useful for teachers to see distribution of exercises' difficulties and potentially identify problematic exercises                                                                                                                         & \G{1}          \\
	Could support dialogue between teachers and parents (e.g., explain how platform diversifies, show students' track record)                                                                                                          & \G{1}          \\
	Construct learning paths by indicating at which mastery level students can switch to another topic                                                                                                                                        & \G{2}          \\

	\midrule
	\multicolumn{2}{@{}l}{\textcolor{focusgroups}{\textbf{\textit{What-if} explanation}}}                                                                                                                                                                         \\
	Highlight progress even when the mastery level stays the same                                                                                                                                                                                        & \G{1}, \G{2}   \\
	To increase motivation, visualise goals and expectations (e.g., thresholds between mastery levels, path towards a goal)                                                                                                             & \G{1}, \G{2}   \\
	Most interesting part because it shows potential progress                                                                                                                                                                                 & \G{1}          \\

	\midrule
	\multicolumn{2}{@{}l}{\textcolor{focusgroups}{\textbf{General comments about explanations}}}                                                                                                                                                                  \\
	Relation between recommendations and explanations is unclear                                                                                                                                                                              & \G{2}          \\
	Orient axes vertically to better represent the idea of ``climbing up''                                                                                                                                                                    & \G{2}          \\
	Good to avoid that students are brainlessly solving exercises; explanations can persuade them to practice attentively                                                                                                                     & \G{2}          \\
	Unclear whether students will really look at the explanations                                                                                                                                                                             & \G{2}          \\
	Potentially only show explanations on demand                                                                                                                                                                                              & \G{2}          \\

	\midrule
	\multicolumn{2}{@{}l}{\textcolor{focusgroups}{\textbf{Control}}}                                                                                                                                                                       \\
	Let students choose non-recommended exercises if they feel their mastery level differs from the system's estimate                                                                                                                  & \G{1}          \\
	Students might lack sufficient self-direction to choose the right topics and exercises                                                                                                                                                    & \G{2}          \\
 \bottomrule
\end{tabular}
\end{table*}

\paragraph{Students Need Goal-Oriented, Motivating Explanations} Participants in \G{2} appreciated that our explanations could stimulate metacognition but at the same time, questioned whether students would pay attention to them. In line with \A{1}--\A{6}, participants deemed our \textit{why} explanation too complex for adolescents and feared the many dots might raise false beliefs that all exercises must be solved instead of a subset. Interestingly, participants in both \G{1} and \G{2} hesitated whether students need to understand the rationale behind recommendations (\textit{why}) as they need to practice anyway. For example, \B{6} said: \quote{Students just want to know which exercises they should solve. Then solve them, period.} Yet, all participants agreed that the main objective in the context of practising should be motivating students and therefore preferred showing the impact of correctly solving exercises (\textit{what-if}). For example, \B{4} stated: \quote{I don't know what the added value is of representing all exercises. I think it's more interesting for [students] to see: if I do this, then that will be the effect on my level.} \B{5} added: \quote{We must avoid students brainlessly completing exercises without understanding why they are making them. Not why the system proposed them, but `what's in it for me', why should I make those exercises.} This relates to another prominent theme: students should especially see their progress and how they can achieve specific goals. For example, \B{7} said: \quote{Students should actually only know why they have a certain mastery level, what steps they should take to get to another level, and which options they have [to get there].} In sum, participants found \textit{what-if} explanations more relevant than \textit{why} explanations for students.

\paragraph{How Appropriate Is Learner Control?} \B{4} alluded to giving students more control over exercises' difficulty instead of fully relying on recommendations: \quote{Can students click on a non-recommended exercise [in the \textit{why} explanation] to try it anyway? Maybe they just guessed five times and therefore have an insufficient level, while they actually do master [the content]. And maybe then they'll say like: `Okay, I'll do my best now'.} Interestingly, \A{5} also suggested this interaction. Participants in \G{2}, however, cautioned that extending such control to freely choosing topics and individual exercises might be counterproductive: they questioned whether young students have sufficient self-direction to make such pedagogically important decisions without guidance.

\paragraph{`Why' Explanations Can Empower Teachers} Whereas participants did not deem \textit{why} explanations suitable for adolescents, they saw potential in them for teachers for four reasons. First, it could educate teachers about how exercises are being recommended: \quote{AI-driven decision-making is still very unfamiliar to many teachers, and this really makes it visual} (\B{3}). Second, teachers could use visualisations like ours to monitor students' progress and exercises' difficulties. The former helps teachers decide which students need personal guidance; the latter supports identifying problematic exercises. Third, the visual \textit{why} explanation could support teachers when talking to parents: \quote{[to tackle questions such as] `Why does my child need to solve these exercises?', this allows them to perfectly explain that [exercise series] are compiled tailored to individual students. [...] It brings nuance to the dialogue, which can sometimes be hard} (\B{3}). Fourth, to stimulate spaced practice, \B{5} proposed an original idea: teachers could manually draw learning paths between different axes in the visualisation, indicating at which mastery level students can switch to another topic. Alternatively, the visualisation could depict automatically generated learning paths, which teachers could tweak. These comments underline that teachers remain important in the context of AI-driven e-learning.

\subsection{Study 4: Think-Aloud Sessions}
\label{sec:prototype3}
\begin{figure*}
  \centering
  \includegraphics[width=.79\linewidth]{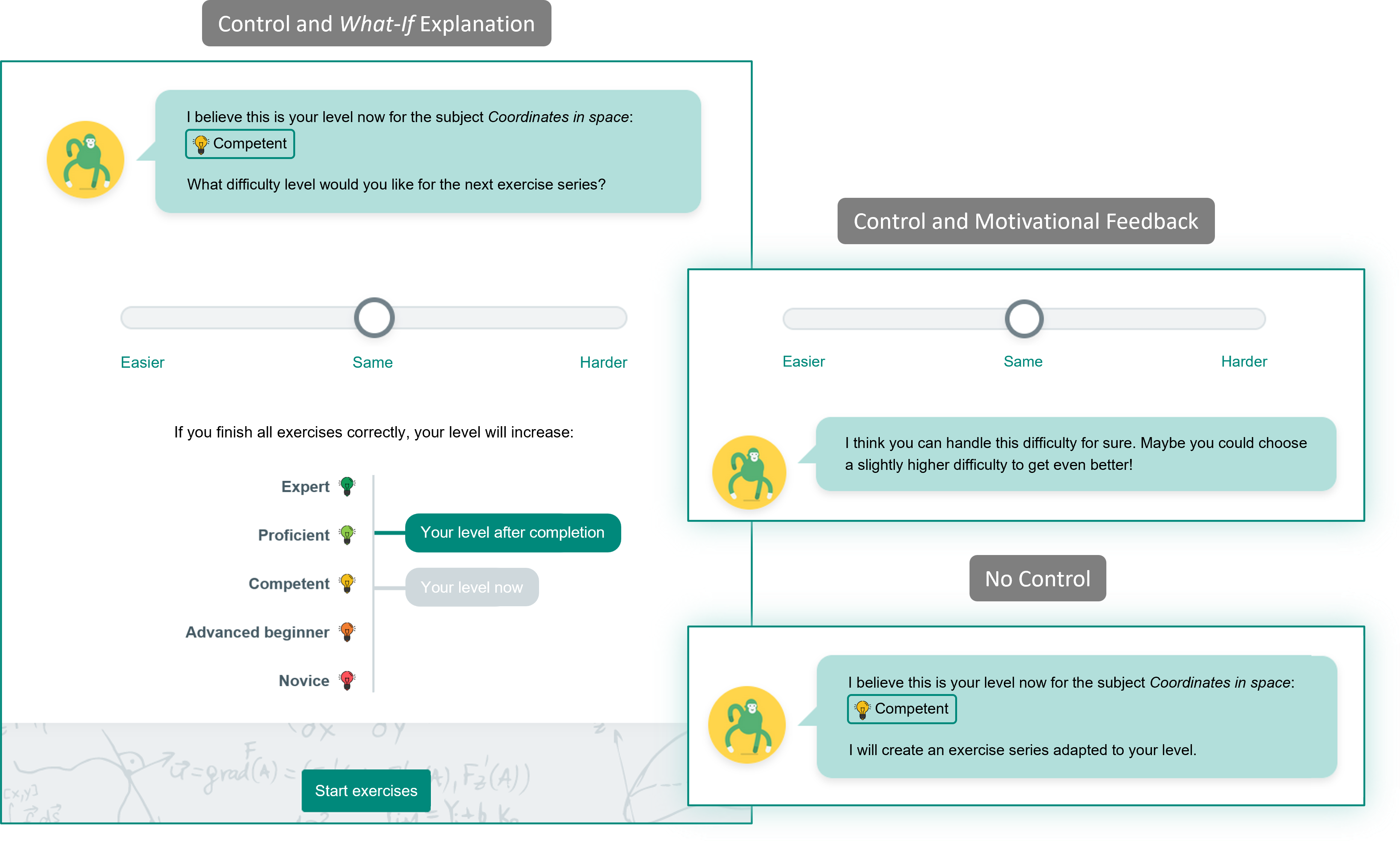}
  \caption{Screenshots of our third prototype with a slider for learner control, linked to a redesigned \textit{what-if} explanation (left) or motivational feedback (right). The latter two components are interactively updated based on the chosen slider value.}
  \Description{Fully described in the caption and main text.}
  \label{fig:prototype3}
\end{figure*}

Based on the feedback in studies 2--3, we made some drastic changes shown in Figure~\ref{fig:prototype3}. Most noticeably, we dropped the \textit{why} explanation. While we could have further improved its design, our studies showed that \textit{why} explanations did not fulfil a primary need for motivation, contrary to \textit{what-if} explanations. In the latter, we oriented the mastery level axis vertically as suggested in \G{2} and discretised it, inspired by the five-stage Dreyfus model~\cite{dreyfus2004fivestage} \revision{and previous work~\cite{ooge2023steering}}. Given the allusions to learner control and better-integrated explanations, we introduced a slider to select the difficulty of recommended exercise series, which updated the \textit{what-if} explanation in real-time according to the handle's position. In addition, we created a design in which the slider is complemented by the dynamic motivational sentences in Table~\ref{tab:wisefeedback}. These sentences were inspired by \textit{wise feedback}, which aims to boost motivation by conveying high standards for learners' performance together with a belief in their potential to reach those standards~\cite{cohen1999mentor,yeager2014breaking}. We opted for this take on wise feedback instead of other motivational techniques such as gamification because it integrates well with the control mechanism, links to learners' mastery level, and requires no further modifications to the e-learning platform.

\begin{table*}
  \caption{Motivational sentences inspired by wise feedback and linked to the slider in Figure~\ref{fig:prototype3}. Whenever the slider handle moves into one of the difficulty subintervals ($0=\text{easier}$, $1=\text{harder}$), a randomly chosen corresponding sentence appears.}
  \label{tab:wisefeedback}
  \footnotesize
\begin{tabular}{@{}ll@{}}
 	\toprule
	\textbf{Difficulty} & \textbf{Motivational Feedback}                                                                                                               \\
	\midrule  
	{[}0.0, 0.2{]}         & I think you can handle exercises that are much harder. I expect a lot from you and am sure you can do it!                            \\
	                    & This level is very low for you. I expect you can solve a more difficult series. I believe in you!                                    \\
	                    & These are very easy exercises. You can surely make them, but I believe you can handle a harder level. Then you'll grow faster! \\

	{[}0.2, 0.4{]}         & I believe you can handle exercises that are more difficult. I believe you can grow even further!                                     \\
	                    & I think you can handle more difficult exercises. That way you will grow faster!                                                    \\
	                    & You can definitely solve easier exercises, but I believe you can handle slightly more difficult exercises. You can do it!            \\

	{[}0.4, 0.6{]}         & I think you can handle this difficulty for sure. Maybe you could choose a slightly higher difficulty to get even better!           \\
	                    & This is a level I believe you can handle. Maybe you can choose a slightly harder level to grow faster.                               \\
	                    & I believe you can solve these exercises, but maybe you could set the difficulty slightly higher. That way you'll get even better!    \\

	{[}0.6, 0.8{]}         & This difficulty is challenging, but the bar is high and I trust in your abilities!                                                   \\
	                    & This is a slightly more challenging level, but I definitely believe you can solve these exercises correctly!                         \\
	                    & I trust that you can solve these difficult exercises. That way, you will also grow faster.                                           \\

	{[}0.8, 1.0{]}         & This difficulty seems very challenging for you. If you think you can handle it, I totally support you!                               \\
	                    & Wow, a challenge! You can always try, I believe in you!                                                                              \\
	                    & This is a very difficult level. But if you think you can handle the exercises, I totally believe in you!                             \\
                     \bottomrule
\end{tabular}
\end{table*}

To evaluate our updated designs, we conducted another round of think-aloud sessions with 6 middle school students (\C{1}--\C{6}, all 7th grade). \revision{Participants explored four interfaces, which either contained only an announcement of the next series, only the slider, or the slider with linked \textit{what-if} explanations and motivational feedback.} Overall, participants did not face major usability issues, but we noticed participants were less fluent \revision{in \anon[LANGUAGE]{Dutch}} than \A{1}--\A{6} and read slowly, so the findings in Table~\ref{tab:thinkaloud2} should be interpreted accordingly. The three lessons below elaborate on our findings.

\begin{table*}
  \centering
  \caption{Main findings of our think-aloud sessions with the prototype in Figure~\ref{fig:prototype3}, ordered by overall theme and frequency.}
  \label{tab:thinkaloud2}
\footnotesize
\begin{tabular}{@{}ll@{}}
	\toprule
	\textbf{Comment}                                                                                  & \textbf{Participants}                    \\

	\midrule
	\multicolumn{2}{@{}l}{\textcolor{thinkaloud2}{\textbf{Control}}}                                                                   \\
	Meaning and functioning of the slider is immediately clear                                        & \C{1}, \C{2}, \C{3}, \C{4}, \C{5}, \C{6} \\
	Not reading or only glancing at the text above the slider                                         & \C{1}, \C{2}, \C{3}, \C{4}, \C{5}, \C{6} \\
	Slider supports reflecting on own mastery level                                                   & \C{2}, \C{3}, \C{4}, \C{5}, \C{6}        \\
	Slider could be more finegrained                                                                  & \C{6}                                    \\

	\midrule
	\multicolumn{2}{@{}l}{\textcolor{thinkaloud2}{\textbf{\textit{What-if} explanation}}}                                                        \\
	Favourite interface (colours, balanced text and visuals, see progress)                            & \C{1}, \C{2}, \C{3}, \C{4}, \C{5}        \\
	Level names `expert', `competent', and `proficient' are unclear, but colours clarify the ordering & \C{1}, \C{2}, \C{3}, \C{5}               \\

	\midrule
	\multicolumn{2}{@{}l}{\textcolor{thinkaloud2}{\textbf{Motivational feedback}}}                                                               \\
	Feedback is motivating and supporting (e.g., to choose harder difficulties)                       & \C{1}, \C{2}, \C{5}, \C{6}               \\
  Feedback is less important than control alone                                                     & \C{3}, \C{4}, \C{6}                      \\
	Feedback is too long                                                                              & \C{2}, \C{3}                             \\
	Feedback supports reflecting on own mastery level                                                 & \C{1}                                    \\
	Harder exercises should not be promoted after errors                                              & \C{4}                                    \\
	\bottomrule
\end{tabular}
\end{table*}

\paragraph{Control Stimulates Self-Reflection} All participants found it intuitive to steer exercises' difficulty with a slider. For example, \C{4} explicitly mentioned: \quote{[I] choose harder [...] because the exercises I had now were a little easy.} Furthermore, \C{2} explained how the slider made them think more about their level: \quote{[without slider] it seems you just have to read the text and start right away and [with slider] you get to choose hard and so on and then you can start.} This demonstrates how our control mechanism encouraged participants to reflect upon which difficulty levels they could handle. For \C{6}, the control mechanism was even the most important feature, underlining what others seemed to concur with: participants highly appreciated learner control.

\paragraph{`What-If' Explanations Clarify Progress} Five participants identified the combination of the slider and \textit{what-if} explanation as their favourite, mainly because the latter showed their progress, was not too textual, and was colourful. For example, \C{4} said: \quote{I can see where I am with my level, and if I solve another exercise, I can also see if I get to another level, or stay at the same. [...] If I'm too low, this can show I need to work my way up. [...] I like that.} In addition, the \textit{what-if} visualisation helped \C{4} understand a misconception: they initially thought that correctly solving a single exercise series would always increase their mastery level by one, for example, from `Competent' to `Proficient.' Not everyone understood these level names, but the accompanying coloured icons clarified their meaning. This illustrates the advantage of visual elements over mere text for linguistically weaker learners. 

\paragraph{Motivational Feedback Can Stimulate Challenging Oneself} Four participants found our take on wise feedback motivating and \C{6} said it stimulated them to choose higher difficulty levels: \quote{I wanted to choose easier, but then I saw [the wise feedback], so I made it a little harder. [I kind of like it] because otherwise you always choose easier.} Yet, participants did not blindly follow the advice: for example, \C{4} once read the motivational feedback but deliberately chose a lower difficulty to avoid more mistakes. In addition, \C{2} and \C{3} found the text rather long and \C{4} raised the question of whether harder exercises should be promoted after incorrect answers. These findings show how motivational feedback can persuade adolescents to reasonably push themselves if they are up for it. However, adolescents might not always read the feedback and prefer feedback tailored to their historical performance.

\subsection{Final Design}
\begin{figure*}
  \centering
  \includegraphics[width=.7\linewidth]{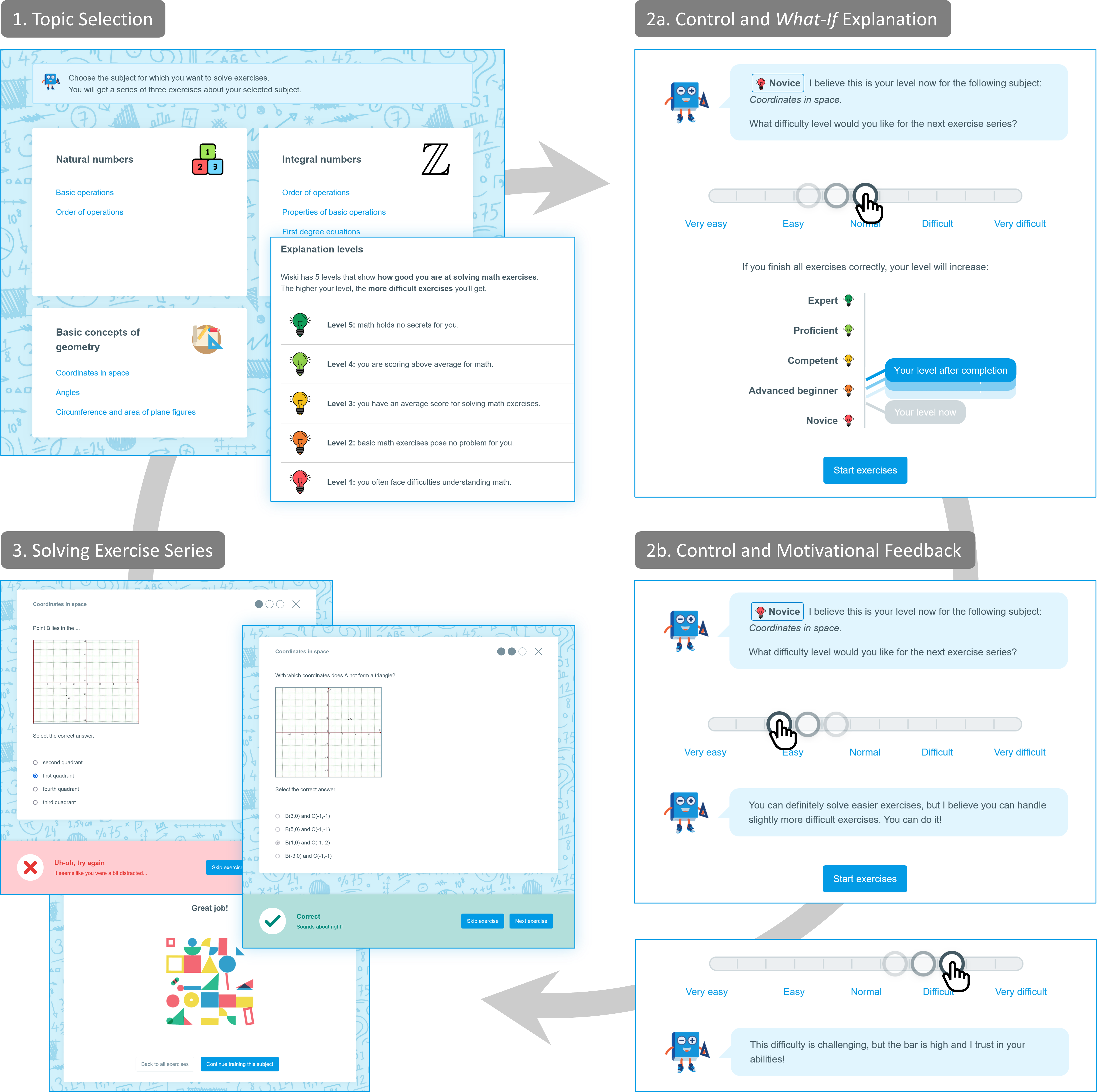}
  \caption[]{Workflow on our e-learning platform: \aptLtoX{\&\#x2776;}{\circledsmall{1}}~Learners choose a topic and inform themselves of the mastery level system; \aptLtoX{\&\#x2777;}{\circledsmall{2}}~The platform compiles a series of three exercises that fit learners' mastery level. Learners can steer its difficulty while seeing a \textit{what-if} explanation or motivational feedback; \aptLtoX{\&\#x2778;}{\circledsmall{3}}~Learners complete the exercises and can restart the practice cycle.}
  \Description{Four groups of screenshots positioned in a cycle, corresponding to the steps in the overall workflow. }
  \label{fig:flow}
\end{figure*}

Figure~\ref{fig:flow} shows our final designs integrated into the workflow of a fully functional e-learning platform with an actual recommendation algorithm. When learners use the slider to steer the difficulty of AI-composed exercise series, they change a difficulty hyperparameter in the recommendation algorithm ($0=\text{very easy}$, $1=\text{very hard}$, steps of 0.1). Thus, learners can adapt the difficulty of exercise series, but not their exact composition, and, contrary to~\cite{ooge2023steering}, learners cannot change their mastery level directly but only steer the platform when they feel it misjudges their mastery level.

\section{Discussion}
Based on the outcomes of our human-centred design process and corresponding user studies, we now propose design implications for visual explanations and learner controls on e-learning platforms.

\subsection{`Why' Explanations for Teachers, Not Adolescents}
\change{While XAI research is booming, it is still unclear which explanation types are suitable for adolescent learners.} Our studies suggest that \textit{why} explanations do not fill pressing needs of adolescents in the context of e-learning platforms that recommend exercises: the adolescents we spoke to did not spontaneously indicate a need to understand the rationale behind recommendations, and only one person mentioned that such understanding could increase their eagerness to solve exercises. As alluded to by teachers and pedagogical experts, this lacking need for explainability might be due to the traditional K-12 school system, which typically imposes tasks on young students and focuses less on self-regulation. Furthermore, our visually demanding \textit{why} explanation sometimes reinforced inaccurate mental models. This could have been due to low visual literacy or heuristic thinking: people resort to intuitive and low-effort thinking to make faster decisions, but this can \change{cause erratic reasoning}~\cite{wang2019designing}. To prevent misunderstanding \change{visual explanations}, \citet{szymanski2021visual} suggested textual annotations, but in our case, these did not bring solace as adolescents often ignored them, potentially because of low reading proficiency \change{or because they did not care about the explanations. Future work could investigate the impact of explanations in a spoken format.}

Yet, \textit{why} explanations might become more prominent for adolescent learners when there is more at stake~\revision{\cite{ooge2022explaining}}. For such cases, we propose two design recommendations based on our think-aloud studies: (a)~adopt visual approaches because graphics and balanced colours seem engaging, and (b)~keep textual annotations concise and linguistically simple to increase the likelihood they are read and understood. Moreover, since our \textit{why} explanation became clearer once adolescents paid closer attention and became more familiar with its visualisation, future studies could introduce complex visual explanations incrementally to foster faster understanding and appreciation. Such explanations should convey learners' mastery level as many of our participants appreciated this information. This underlines the relevance of research into open learner models~\cite{bull2020there,bull1995did,bull2004open,bull2007student,mabbott2006student}. For teachers, our focus groups indicated that \textit{why} explanations could readily be of great benefit when integrating e-learning platforms into their teaching: \textit{why} explanations could help them monitor learners, understand and steer recommendations, and improve communication with parents. \revision{This aligns with goals in the learning analytics community such as enhancing learner engagement, transparency, and control in AI-supported systems}~\cite{bodily2018open,bodily2017review}.

\subsection{`What-If' Explanations and Learner Control to Motivate and Engage}
As all educational stakeholders participating in our studies were strongly concerned with motivating students, we focused our designs on \textit{what-if} explanations, and as such explored a potential bridge between XAI and motivation, similar to~\cite{conijn2023effects}. Our findings suggest that \textit{what-if} explanations could foster motivation: adolescents and pedagogical experts praised them for showing the potential positive impact of practice on mastery level and instilling goals to work towards. \revision{This resonates with motivation theory on mastery-approach goal orientation, where learners are focused on the intrinsic value of learning and try to improve their competence level based on self-referenced standards~\cite{ames1992classrooms}.} To increase motivation through self-efficacy, we also proposed motivational sentences inspired by wise feedback, which seemed to stimulate adolescents to challenge themselves more.

Furthermore, while our designs initially focused on transparency through explanations, both adolescents and teachers raised a need for learner control over recommended exercises. We therefore allowed learners to manipulate the difficulty of AI-composed exercise series; a rarely studied approach~\cite{brusilovsky2023ai}, which was well received and proved to be an adequate level of control for adolescents. These findings align with previous work that showed learners appreciate control over their learning process~\cite{ooge2023steering,vandewaetere2011can,clark2011elearning}. Yet, as \citet{clark2011elearning} pointed out, learner control might not always be appropriate in early learning phases. Similar to how we observed inattentiveness harmed mental models about how exercises were recommended, carelessly executing control over which exercises to practice could harm learning.

Even though adolescents in our studies reported appreciating \textit{what-if} explanations, we observed they often ignored explanations or did not link them to the corresponding recommendations when both were disconnected and the explanations were static. This confirms previous findings on people often not cognitively engaging with explainability tools~\cite{liao2022humancentered,miller2023explainable}. We hypothesised the static character of our explanations contributed to this low cognitive engagement and therefore linked them interactively to our control mechanism. This decision was also justified based on earlier research showing that communicating the impact of learner control is essential to foster trust in e-learning platforms~\cite{ooge2023steering}.

To further stimulate cognitive engagement and self-reflection in learners, our \textit{what-if} explanations could be extended by showing not only potential improvement in learning in the best case but also potential decline in the worst case or the expected change based on learning analytics of similar learners. This setup would allow to explore the trade-off between stimulating motivation while staying realistic: for example, \textit{what-if} explanations that show large progress could lose their initial motivating effect once it becomes clear they set impossible goals. Similarly, the motivational sentences we linked to the control mechanism could be made adaptive so they only push learners towards harder exercises when appropriate.

\subsection{Limitations and Future Work}
Our research has several limitations which restrict how well our findings generalise. Regarding design, we based the phrasings of our motivational sentences on existing literature about wise feedback~\cite{yeager2014breaking} but did not consult pedagogical experts. Regarding participants, we only involved adolescents from 7th to 11th grade in our study. While this is a strong contribution as adolescents are often overlooked in XAI research, \revision{learners of other ages might have different needs regarding explainability and learning agency, influenced, for example, by their reading and self-regulation skills. Future studies could compare different age groups explicitly.} Regarding methods, we measured cognitive engagement only through observation. Future studies could use eye-tracking to check whether and how adolescents analyse explanations and control mechanisms. \revision{Finally, follow-up studies on deployed e-learning platforms should verify whether our preliminary findings hold in larger samples during real learning activities. Large-scale randomised controlled experiments could quantitatively assess how control mechanisms combined with \textit{what-if} explanations and motivational feedback affect metacognition, motivation, trust, learning performance, and chosen difficulty levels.}

\section{Conclusion}
E-learning platforms for adolescents are abundant, but \change{studies} on designing effective explanations and learner control mechanisms for them are not. We derived many lessons regarding these aspects from a four-stage design process involving adolescents, EdTech professionals, teachers, and pedagogical experts. In particular, we found that \textit{why} explanations do not necessarily fulfil explainability needs for young learners but could be useful for teachers. Furthermore, \textit{what-if} explanations and supportive sentences based on wise feedback were deemed motivating, and linking these to a control mechanism seemed to stimulate cognitive engagement. These findings contribute to a better understanding of how young learners best steer AI-supported e-learning systems and inspire follow-up randomised controlled experiments.

\begin{acks}
This work was funded by the Research Foundation Flanders (grants G067721N and V431323N), KU Leuven (grant C14/21/072), and Flanders Innovation \& Entrepreneurship (imec-icon AI-Driven e-Assessment project). Joke Vandepitte recruited students and teachers. Bram Faems recruited pedagogical experts and co-conducted the focus groups.
\end{acks}

\bibliographystyle{ACM-Reference-Format}
\bibliography{referencesLAK2025}

\end{document}